\documentclass[epj]{svjour}
\usepackage{amsmath}
\usepackage{graphics,color,subfigure}
\usepackage{epsfig}
\usepackage{dcolumn}
\usepackage{bm}
\usepackage{longtable}

\def\gamf{\gamma_5}
\def\la{\langle}
\def\ra{\rangle}

\def\chip{\chi_+}

\def\psib{\bar{\psi}}

\def\Nb{\bar{N}}

\newcommand{\Tb}{\bar{T}}

\usepackage[colorlinks=true,
                  urlcolor=blue,
                  linkcolor=blue,
                  citecolor=blue,
                  bookmarks=true,
                  bookmarksopen=true,
                  bookmarksnumbered=true]{hyperref}
\begin{document}
\title{Relations for low-energy constants in baryon chiral perturbation theory with explicit $\Delta(1232)$ derived from the chiral quark model}
\author{Jun Jiang\inst{1},
Shao-Zhou Jiang\inst{2}\thanks{e-mail: jsz@gxu.edu.cn},
Shi-Yuan Li\inst{1},
Yan-Rui Liu\inst{1}\thanks{e-mail: yrliu@sdu.edu.cn},
Zong-Guo Si\inst{1},
Hong-Qian Wang\inst{1}
}                     
%
%
\institute{School of Physics, Shandong University, Jinan, Shandong 250100, China
\and Guangxi Key Laboratory for Relativistic Astrophysics, School of Physical Science and Technology, Guangxi University, Nanning, Guangxi 530004, China}
\date{Received: date / Revised version: date}
%
\abstract{
We study the relations between low-energy constants in the chiral Lagrangians with $\Delta(1232)$ and those in the quark-level description model up to the third chiral order. Ten structure correspondences are involved in getting the relations. This situation is more complicated than the spin-1/2 baryon case. The obtained results may help to further investigations involving the $\Delta(1232)$ baryons.
%
} 
\titlerunning{xxx}
\authorrunning{J. Jiang et al.}
\maketitle
%
\section{Introduction}\label{sec1}

In the baryon chiral perturbation theory ($\chi$PT), the contributions of higher resonances are implicitly included in the low-energy constants (LECs) of the Lagrangian. Such resonances can also be considered as explicit degrees of freedom in the language of effective field theory, which introduces additional scales. In the two-flavor case, the mass difference between $\Delta$ (the lowest $\pi N$ resonance) and $N$ is $\Delta\equiv m_\Delta-m_N\approx294$ MeV and it goes to zero in the large $N_c$ limit \cite{tHooft:1973alw,Witten:1979kh}. It is also known that the coupling between $\Delta$, nucleon, and pion is strong.  Therefore, when the applicability region of the baryon $\chi$PT is extended, the $\Delta$ baryon should be treated as an explicit field. Here we use $\Delta\chi$PT to indicate such an effective theory. In the conventional baryon $\chi$PT, $m_N$ is close to the scale of chiral symmetry breaking ($\Lambda_\chi$) and the power counting problem is complicated because the momenta of hadrons should be much smaller than $\Lambda_\chi$ for an ideal chiral theory. In $\Delta\chi PT$, the new scale $m_\Delta$ makes the power counting problem more complicated than the baryon $\chi$PT. A widely adopted power counting scheme is the small scale expansion (the mass difference $\Delta$ and the soft momentum $p$ belong to the same order) \cite{Hemmert:1997ye}, but the $\delta=\Delta/\Lambda_\chi\sim m_\pi/\Delta$ expansion \cite{Pascalutsa:2002pi,Pascalutsa:2005ts} was also used. The following discussions  will focus on the $SU(2)$ case and the small scale expansion is involved.

The extension of $\chi$PT to $\Delta\chi$PT introduces more chirally invariant coupling terms and corresponding coupling constants. Now there are three sectors of baryon Lagrangians: ${\cal L}_{\pi NN}$, ${\cal L}_{\pi\Delta\Delta}$, ${\cal L}_{\pi N\Delta}$. All the interaction terms up to the fourth order in the small scale expansion have been constructed  \cite{Fettes:2000gb,Jiang:2017yda,Jiang:2018mzd}. It is a problem to calculate the related LECs from the underlying theory. Up to now,  various theoretical methods to determine or constrain their values have been adopted, e.g.  the resonance saturation estimation \cite{Pich:2008xj,Ecker:1988te,Bernard:1993fp,Liu:2010bw,Liu:2011mi,Du:2016tgp}.

Since there are lots of experimental data about nucleons, one usually extracts the related LECs in ${\cal L}_{\pi NN}$ from measurements. For the unstable $\Delta(1232)$, available experimental data to determine the relevant LECs are relatively scarce. If one can find some relations between LECs in ${\cal L}_{\pi\Delta\Delta}$ and ${\cal L}_{\pi N\Delta}$ to those in ${\cal L}_{\pi NN}$ with some symmetries or models, the results are also helpful to the application of $\Delta\chi$PT.  The large $N_c$ expansion and quark model can be adopted for such a purpose. Previously in Ref. \cite{Jiang:2022gjy}, we analyzed the LEC relations between the  $SU(2)$ baryon $\chi$PT and the $SU(3)$ baryon $\chi$PT through the chiral quark model up to the third chiral order. In the present paper, we explore similarly the LEC relations in $\Delta\chi$PT up to the third chiral order.

In Ref. \cite{Jiang:2022gjy}, the involved matter fields are spin-1/2 baryons and quarks. In the present case, there are different features because we use the Rarita-Schwinger (RS) field \cite{Rarita:1941mf} to describe the spin-3/2 $\Delta(1232)$.  In principle, each hadron-level interaction vertex in $\Delta\chi$PT should have a corresponding quark-level description. Unlike the spin-1/2 case where the structure correspondence between $\pi NN$ terms and $\pi qq$ terms is one-to-one \cite{Jiang:2022gjy}, the complicated Lorentz structures for the $\Delta$ Lagrangians and the existent redundant degrees of freedom for RS fields make the correspondence ambiguous. We explore this issue in the present study. The procedure to get the LEC relations is similar to the one adopted in Ref. \cite{Jiang:2022gjy}. Now it is:
\begin{center}
\includegraphics{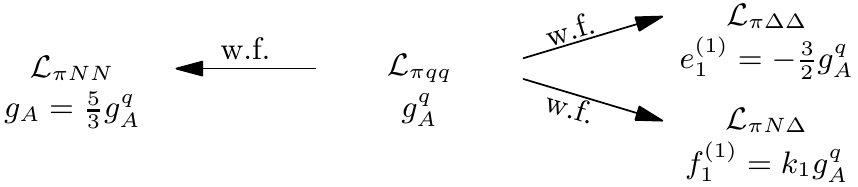}
\end{center}
where w.f. means quark wave function of $N$ or $\Delta$, $k_1$ is a constant, and $g_A$, $g_A^q$, $e_1^{(1)}$, and $f_1^{(1)}$ are the coupling constants in the leading order ${\cal L}_{\pi NN}$, ${\cal L}_{\pi qq}$, ${\cal L}_{\pi \Delta\Delta}$, and ${\cal L}_{\pi N\Delta}$, respectively (see Table \ref{p1}). We mainly concentrate on the coupling constant relations between ${\cal L}_{\pi \Delta\Delta, \pi N\Delta}$ and ${\cal L}_{\pi qq}$. Since ${\cal L}_{\pi qq}$ in the quark model we use is treated as the quark-level description of ${\cal L}_{\pi NN}$, ${\cal L}_{\pi\Delta\Delta}$, and ${\cal L}_{\pi N\Delta}$, we also call the quark-level coupling constants LECs.

To find the structure correspondences in the present study, we first reorganize terms in ${\cal L}_{\pi\Delta\Delta}$ and ${\cal L}_{\pi N\Delta}$. Then the necessary operators and relevant properties are given. From the features of the interation terms up to the third chiral order, one sets ten structure correspondences. Considering possible inequivalence when comparing hadron- and quark-level  descriptions, we introduce several multiplication factors in the correspondences. Then the investigated relations are not difficult to obtain.

The present paper is organized as follows. We collect the needed Lagrangians in Sec. \ref{sec2}. The quark wave functions of baryons and the properties for the RS field are given in Sec. \ref{sec3}. In Sec. \ref{sec4}, we set the structure correspondences and list the obtained LEC relations. In Sec. \ref{sec5}, we explore further the correspondences and the introduced multiplication factors. The final Sec. \ref{sec6} present some discussions.

\section{Chiral Lagrangians in hadron- and quark-level descriptions}\label{sec2}

Both baryons and quarks are involved in the present work. The spin-1/2 fields are
\begin{eqnarray}\label{defpsi}
N=\left(\begin{array}{c}p\\n\end{array}\right),\qquad
\psi_q=\left(\begin{array}{c}u\\d\end{array}\right).
\end{eqnarray}
For the spin-3/2 $\Delta$ fields, one may choose different representations in flavor space. In Ref. \cite{Jiang:2017yda}, the Lagrangians with the form of isovector-isospinor RS field were given. In Ref. \cite{Jiang:2018mzd}, those with the form of totally symmetrical tensor were presented. One can find the relations between these two forms of Lagrangians in the latter reference. The quark content is manifest in the latter form and it is convenient for our purpose. Therefore, we choose such a representation, i.e., the $\Delta$ fields in flavor space are denoted by  $T^{abc}$ with
\begin{eqnarray}\label{T111}
	&T^{111}=\Delta^{++},\quad T^{112}=T^{121}=T^{211}=\frac{\Delta^+}{\sqrt3},&\nonumber\\
&T^{122}=T^{212}=T^{221}=\frac{\Delta^0}{\sqrt3}, \quad T^{222}=\Delta^-.&
\end{eqnarray}

Other necessary definitions in the Lagrangians are
\begin{eqnarray}
u&=&\exp(\frac{i\phi}{2f_\pi}), \quad \phi=\pi^i\tau^i,\nonumber\\
r^\mu&=&v^\mu+a^\mu,\quad l^\mu=v^\mu-a^\mu,\nonumber\\
u^\mu&=&i\{u^\dag(\partial^\mu-ir^\mu)u-u(\partial^\mu-il^\mu)u^\dag\},\nonumber\\
\chi&=&2B_0(s+ip),\quad \chi_{\pm}=u^\dag\chi u^\dag\pm u\chi^\dag u,\nonumber\\
h^{\mu\nu}&=&\nabla^\mu u^\nu+\nabla^\nu u^\mu,\nonumber\\
F_L^{\mu\nu}&=&\partial^{\mu}l^{\nu}-\partial^{\nu}l^{\mu}-i[l^{\mu},l^{\nu}],\nonumber\\
F_R^{\mu\nu}&=&\partial^{\mu}r^{\nu}-\partial^{\nu}r^{\mu}-i[r^{\mu},r^{\nu}],\nonumber\\
f_+^{\mu\nu}&=&u F_L^{\mu\nu} u^\dag+u^\dag F_R^{\mu\nu} u,\nonumber\\
f_-^{\mu\nu}&=&u F_L^{\mu\nu} u^\dag-u^\dag F_R^{\mu\nu} u=-\nabla^\mu u^\nu+\nabla^\nu u^\mu,\nonumber\\
\Gamma^\mu&=&\frac12\{u^\dag(\partial^\mu-ir^\mu)u+u(\partial^\mu-il^\mu)u^\dag\},\nonumber\\
\chi^\mu_\pm&=&u^\dag\tilde{\nabla}^\mu\chi u^\dag \pm u\tilde{\nabla}^\mu\chi^\dag u =\nabla^\mu\chi_\pm-\frac{i}{2}\{\chi_{\mp},u^\mu\},\nonumber\\
\nabla^\mu O&=&\partial^\mu O+[\Gamma^\mu,O],\nonumber\\
\tilde{\nabla}^\mu\chi&=&\partial^\mu\chi-ir^\mu\chi+i\chi l^\mu,\nonumber\\
D^\mu \psi&=&\partial^\mu\psi+\Gamma^\mu\psi,\nonumber\\
D^\mu T_{abc}&=&\partial^\mu T_{abc}+{\Gamma_a}^{d,\mu}T_{dbc}+{\Gamma_b}^{d,\mu}T_{adc}+{\Gamma_c}^{d,\mu}T_{abd},\nonumber\\
D_{\nu\lambda\rho\cdots}&=&D_\nu D_\lambda D_\rho\cdots+\mathrm{full}\,\mathrm{permutation}\,\mathrm{of }\, D\mathrm{'s}.
\end{eqnarray}
Here, $\tau^i$ $(i=1,2,3)$ are the Pauli matrices and $B_0$ is a constant related to the quark condensate. $\psi$ represents $N$ or $\psi_q$ in Eq. (\ref{defpsi}) and ${\cal O}$ denotes the building block $u^\mu$, $\chi_\pm$, $h^{\mu\nu}$, or $f^{\mu\nu}_\pm$. Of the scalar ($s$), pseudoscalar ($p$), vector ($v_\mu$), and axial-vector ($a_\mu$) external sources, the last one is  usually taken as traceless. Different from Ref. \cite{Jiang:2022gjy}, we have $\la v^\mu\ra\neq0$ and thus $\la f_+\ra\neq0$ in the present $SU(2)$ case. For convenience \cite{Fettes:2000gb}, we also separate the matrix $X=\chi_\pm$, $\chi^\mu_\pm$,  or $f_+^{\mu\nu}$ into the trace part $\langle X\rangle$ and the traceless part $\tilde{X}=X-\frac12\langle X\rangle$.

In Ref. \cite{Jiang:2022gjy}, we have classified the fermion interaction terms into different groups according to the Lorentz and flavor structures. In the present case, the classification is slightly different because of the properties of spin-3/2 fields. We show the results in Tables \ref{p1}-\ref{p3}. For the RS field $T_\mu$, one has to use subsidiary conditions to eliminate its redundant degrees of freedom. Therefore, the relations like
\begin{eqnarray}
&&\gamma^\mu\gamma_5T^\nu-\gamma^\nu\gamma_5T^\mu\doteq i\varepsilon^{\mu\nu\lambda\rho}\gamma_\rho T_\lambda\nonumber\\
&\doteq& -\frac{1}{m_\Delta}
\varepsilon^{\mu\nu\lambda\rho}D_\rho T_\lambda\quad (\varepsilon^{0123}=-1)
\end{eqnarray}
connect different Lorentz structures. The one-to-one structure correspondences between hadron- and quark-level descriptions similar to those in Refs. \cite{Drechsel:1983hny,Jiang:2022gjy} do not exist.

\begin{table*}[htbp]\centering
\caption{The ${\cal O}(p^1)$ chiral-invariant terms of the Lagrangians ${\cal L}_{\pi NN}$, ${\cal L}_{\pi qq}$, ${\cal L}_{\pi\Delta\Delta}$, and ${\cal L}_{\pi N\Delta}$. Their corresponding coupling constants are $g_A$, $g_A^q$, $e_i^{(1)}$, and $f_i^{(1)}$ , respectively. $\psi$ in ${\cal L}_{\pi NN/\pi qq}$ means $N/\psi_q$ of Eq. (\ref{defpsi}).}\label{p1}
\begin{tabular}{cc|cc|cc}\hline\hline
\multicolumn{2}{c|}{${\cal L}_{\pi NN/\pi qq}$}&\multicolumn{2}{c|}{${\cal L}_{\pi\Delta\Delta}$}&\multicolumn{2}{c}{${\cal L}_{\pi N\Delta}$}\\	
$i$&$g_A/g_A^q$&$i$&$e_i^{(1)}$  &  $i$ & $f_i^{(1)}$   \\
1&$\frac12\psib u^{\mu}\gamma_\mu\gamma_5\psi$ &1 &$\Tb^{ a b c,\lambda}u^{ad,\mu}\gamma_{\mu}\gamma_5T^{ b c d}_\lambda$
&1 & $\epsilon^{ a b}\Nb^c u^{a d,\mu}T^{b c d}_\mu+h.c.$\\
\hline\hline
\end{tabular}
\end{table*}

\begin{table*}[htbp]\centering
\caption{The ${\cal O}(p^2)$ chiral-invariant Lagrangians ${\cal L}_{\pi NN}$, ${\cal L}_{\pi qq}$, ${\cal L}_{\pi\Delta\Delta}$, and ${\cal L}_{\pi N\Delta}$. $\psi$ in ${\cal L}_{\pi NN/\pi qq}$ means $N/\psi_q$ of Eq. (\ref{defpsi}) and $\langle\cdots\rangle$ means trace in flavor space. The symbol $[\cdots]_+$ stands for $[\cdots]+\mathrm{h.c.}$. We use $i$ to label the interaction terms and $\alpha_i^{(2)}$, $\beta_i^{(2)}$, $\hat{e}_i^{(2)}$, and $\hat{f}_i^{(2)}$ to denote their coupling constants, respectively. The meaning of symbols like $f^{(2)}_{1-2-2}$ is explained in the context.}\label{p2}
\begin{tabular}{cc|cc|cc}\hline\hline
\multicolumn{2}{c|}{${\cal L}_{\pi NN/\pi qq}$}&\multicolumn{2}{c|}{${\cal L}_{\pi\Delta\Delta}$}&\multicolumn{2}{c}{${\cal L}_{\pi N\Delta}$}\\
$i$&$\alpha_{i}^{(2)}/\beta_{i}^{(2)}$&$i$&$\hat{e}_i^{(2)}$ &		$i$ & $\hat{f}_i^{(2)}$  \\
1 & $\psib\la u^{\mu}u_{\mu}\ra\psi$ &1 &$e^{(2)}_5: \Tb^{ a b c,\lambda}\langle u^{\mu} u_{\mu}\rangle T^{abc}_{\lambda}$ && \\
 &  &2 &$e^{(2)}_2: \Tb^{abc,\lambda}u^{ag,\mu}u^{be}_{\mu}T^{cge}_{\lambda}$ & \\
 &  &3 & $e^{(2)}_{3+4}: \Tb^{abc}_\mu\langle u^\mu u^\nu\rangle T^{abc}_\nu$ &\\
 &  & 4&$e^{(2)}_1: \Tb^{ abc}_\mu u^{ag,\mu}u^{be,\nu}T^{cge}_\nu$&1&
$f^{(2)}_{1-2-2}: [\epsilon^{bc}\Nb^a u^{ag,\mu} u^{be,\nu} (\gamma_\mu\gamf T^{cge}_\nu+\gamma_\nu\gamf T^{cge}_\mu)]_+$\\
\hline
2 & $i \psib u^{\mu}u^{\nu}\sigma_{\mu\nu}\psi$& 5 &
       $e^{(2)}_{3-4}: i\Tb^{abc,\lambda} (u^{\mu}u^\nu)^{ae}\sigma_{\mu\nu} T^{bce}_\lambda$ &2 &$f^{(2)}_1: [\epsilon^{ab}\Nb^c(u^\mu u^\nu)^{ae}(\gamma_\mu\gamf T^{bce}_\nu-\gamma_\nu\gamf T^{bce}_\mu)]_+$  \\
  &                                  &&&&\\
 \hline
3 & $\psib\la u^{\mu}u^{\nu}\ra D_{\mu\nu}\psi$&6 &	$e^{(2)}_7: \Tb^{ a b c,\lambda}\langle u^{\mu}u^{\nu}\rangle D_{\mu\nu}T^{abc}_{\lambda}$  &&\\
		&                                      &7 &$e^{(2)}_6: \Tb^{ a b c,\lambda}u^{ag,\mu}u^{be,\nu}D_{\mu\nu}T^{cge}_{\lambda}$\\\hline
4 & $\psib \tilde{f}_{+}^{\mu\nu}\sigma_{\mu\nu}\psi$ & 8&$e^{(2)}_9:\Tb^{ a b c,\lambda} (\tilde{f}_{+ }^{\mu\nu})^{ae}\sigma_{\mu\nu} T^{bce}_\lambda$&3&$f^{(2)}_3: [i\epsilon^{ a b}\Nb^c(\tilde{f}_+^{\mu\nu})^{ae}(\gamma_{\mu}\gamf T^{bce}_\nu-\gamma_{\nu}\gamf T^{bce}_\mu)]_+$ \\
5 & $\psib\la f_{+}^{\mu\nu}\ra\sigma_{\mu\nu}\psi$& 9&	$e^{(2)}_8: \Tb^{ a b c,\lambda} \langle f_+^{\mu\nu}\rangle\sigma_{\mu\nu} T^{abc}_\lambda$  &&\\\hline
6 & $\psib\tilde{\chi}_+\psi$  &10&	$e^{(2)}_{11}: \Tb^{ a b c,\lambda}\tilde{\chi}_+^{ae}T^{bce}_\lambda$&&\\
7 & $\psib\la\chip\ra\psi$ &11&$e^{(2)}_{10}: \Tb^{ a b c,\lambda}\langle\chi_+\rangle T^{abc}_\lambda$&\\
\hline\hline
\end{tabular}
\end{table*}

\begin{table*}[htbp]\centering
\caption{The ${\cal O}(p^3)$ chiral-invariant Lagrangians ${\cal L}_{\pi NN}$, ${\cal L}_{\pi qq}$, ${\cal L}_{\pi\Delta\Delta}$, and ${\cal L}_{\pi N\Delta}$. $\psi$ in ${\cal L}_{\pi NN/\pi qq}$ means $N/\psi_q$ of Eq. (\ref{defpsi}) and $\langle\cdots\rangle$ means trace in flavor space. The symbol $[\cdots]_+$ stands for $[\cdots]+\mathrm{h.c.}$. We use $i$ to label the interaction terms and $\alpha_i^{(3)}$, $\beta_i^{(3)}$, $\hat{e}_i^{(3)}$, and $\hat{f}_i^{(3)}$ to denote their coupling constants, respectively. The meaning of symbols like $e^{(3)}_{2-7-3+4}$ is explained in the context.}\label{p3}
\begin{tabular}{cc|cc|cc}\hline\hline
\multicolumn{2}{c|}{${\cal L}_{\pi NN/\pi qq}$}&\multicolumn{2}{c|}{${\cal L}_{\pi\Delta\Delta}$}&\multicolumn{2}{c}{${\cal L}_{\pi N\Delta}$}\\
$i$&$\alpha_{i}^{(3)}/\beta_{i}^{(3)}$&$i$&$\hat{e}_i^{(3)}$  &		$i$ & $\hat{f}_i^{(3)}$   \\
	
	1 & $\psib\la u^{\mu}u_{\mu}\ra u^{\nu}\gamma_{\nu}\gamf \psi$ & 1 &$e^{(3)}_8: \Tb^{ a b c,\lambda}\langle u^\mu u_\mu\rangle u^{ae,\nu}\gamma_{\nu}\gamf T^{bce}_{\lambda}$ &1 &$f^{(3)}_1: [\epsilon^{bc}\Nb^d \langle u^\mu u_\mu\rangle u^{be,\nu}T^{cde}_\nu]_+$ \\
	2 & $\psib\la u^{\mu}u^{\nu}\ra u_{\mu} \gamma_{\nu}\gamf\psi$ &  2&$e^{(3)}_6: \Tb^{ a b c,\lambda}\langle u^\mu u^\nu\rangle u^{ae}_\mu\gamma_\nu\gamf T^{bce}_\lambda$ &2 &$f^{(3)}_{1+3}: [\epsilon^{  b c}\Nb^d\langle u^\mu u^{\nu}\rangle u^{be}_\mu T^{cde}_{\nu}]_+$\\
	3 & $\varepsilon_{\mu\nu\lambda\rho}\psib\la u^{\mu}u^{\nu}u^{\lambda}\ra D^{\rho}\psi$& 3 &$e^{(3)}_{2-7-3+4}:\varepsilon_{\mu\nu\lambda\rho}\Tb^{abc,\eta}\langle u^\mu u^\nu u^\lambda\rangle D^\rho T^{abc}_\eta$& & \\
	&& 4 &$e^{(3)}_5: \Tb^{ a b c,\lambda}u^{ad,\mu}u^{be}_{\mu}u^{cf,\nu}\gamma_{\nu}\gamf T^{def}_\lambda$&3&$f^{(3)}_2: [\epsilon^{  b c}\Nb^d(u^\mu u^\nu)^{be}u^{df}_{\mu}T^{cef}_\nu]_+$\\
&& 5 &$e^{(3)}_1: \Tb^{ a b c}_\rho u^{ad,\rho}u^{be,\nu}u^{cf,\lambda}\gamma_{\nu}\gamf T^{def}_{\lambda}$&4&$f^{(3)}_7: [i\epsilon^{ bc}\Nb^d(u^\mu u^\nu)^{be}u^{df,\lambda}\sigma_{\mu\nu}T^{cef}_\lambda]_+$   \\
&& 6 &$e^{(3)}_{2+7}: \Tb^{abc}_\rho\langle u^\rho u^\lambda\rangle u^{ae,\nu}\gamma_\nu\gamf T^{bce}_\lambda$& 5&$f^{(3)}_8: [i\epsilon^{  b c}\Nb^d\langle u^\mu u^\lambda\rangle u^{be,\nu}\sigma_{\mu\nu}T^{cde}_\lambda]_+$  \\
&& 7 &$e^{(3)}_{3+4}: [\Tb^{abc}_\rho\langle u^\rho u^\nu\rangle u^{ae,\lambda}\gamma_\nu\gamf T^{bce}_\lambda]_+$&   \\
	&&8&$e^{(3)}_{2-7}: \Tb^{abc}_\rho[u^\rho,u^\lambda]^{ad}u^{be,\nu}\gamma_\nu\gamf T^{cde}_\lambda$ & \\
\hline
	4 & $\psib\la u^{\mu}u^{\nu}\ra u^{\lambda} \gamma_{\mu}\gamf D_{\nu\lambda}\psi$ & 9 &$e^{(3)}_{10}: \Tb^{ a b c,\rho}\langle u^{\mu}u^\nu\rangle u^{ae,\lambda}\gamma_{\mu}\gamf D_{\nu\lambda}T^{bce}_\rho$&6&$f^{(3)}_{5+6}: [\epsilon^{b c}\Nb^d\langle u^\mu u^\nu\rangle u^{be,\lambda}D_{\nu\lambda}T^{cde}_\mu]_+$ \\
	5 & $\psib\la u^{\mu}u^{\nu}\ra u^{\lambda} \gamma_{\lambda}\gamf D_{\mu\nu}\psi$ &10& $e^{(3)}_{11}: \Tb^{ a b c,\rho}\langle u^\mu u^\nu\rangle u^{ae,\lambda}\gamma_{\lambda}\gamf D_{\mu\nu}T^{bce}_\rho$ &7&$f^{(3)}_5: [\epsilon^{  b c}\Nb^d\langle u^\mu u^\nu\rangle u^{be,\lambda}D_{\mu\nu}T^{cde}_\lambda]_+$ \\
	&&11 &$e^{(3)}_9: \Tb^{ a b c,\rho}u^{ad,\mu}u^{be,\nu}u^{cf,\lambda}\gamma_{\mu}\gamf D_{\nu\lambda}T^{def}_\rho$ &8&$f^{(3)}_4: [\epsilon^{  b c}\Nb^d(u^\mu u^\nu)^{be}u^{df,\lambda}D_{\mu\lambda}T^{cef}_\nu]_+$\\
\hline
	6 & $[\psib u_{\mu}h^{\mu\nu}D_{\nu}\psi]_+$ &12 &$e^{(3)}_{19}: [\Tb^{ a b c,\rho}(u_\mu h^{\mu\nu})^{ae}D_\nu T^{bce}_{\rho}]_+$ & &\\
	7 & $[\psib u^{\mu}h^{\nu\lambda}D_{\mu\nu\lambda}\psi]_+$ &13 &$e^{(3)}_{20}: [\Tb^{ a b c,\rho}(u^\mu h^{\nu\lambda})^{ae}D_{\mu\nu\lambda}T^{bce}_\rho]_+$ && \\
	8 & $i \psib\la u^{\mu}h^{\nu\lambda}\ra\sigma_{\mu\nu}D_{\lambda}\psi$ &14&$i\bar{T}^{abc,\rho}\la u^{\mu}h^{\nu\lambda}\ra\sigma_{\mu\nu}D_{\lambda}T^{abc}_\rho$& &\\
	&& &&9&$[\epsilon^{  b c}\Nb^d(u^\mu h^{\nu\lambda})^{be} D_\lambda (\gamma_{\mu}\gamf T^{cde}_\nu+\gamma_{\nu}\gamf T^{cde}_\mu)]_+$\\
         &&&&10&$f^{(3)}_{16}: [\epsilon^{  b c}\Nb^du^{be,\mu}h^{df,\nu\lambda}\gamma_{\mu}\gamf D_\lambda T^{cef}_\nu]_+$\\\hline
	9 & $i \psib [\tilde{f}_+^{\mu\nu},u_{\mu}] \gamma_{\nu}\gamf \psi$ &15&$e^{(3)}_{27}: i\Tb^{ a b c,\rho}[\tilde{f}_+^{\mu\nu},u_{\mu}]^{ae}\gamma_\nu\gamf T^{bce}_\rho$ &11&$f^{(3)}_{21}: [i\epsilon^{  b c}\Nb^d[\tilde{f}_+^{\mu\nu},u_\mu]^{be} T^{cde}_\nu]_+$\\
	10 & $i \varepsilon_{\mu\nu\lambda\rho}\psib\la f_{+}^{\mu\nu}\ra u^{\lambda}D^{\rho}\psi$ &16 &$e^{(3)}_{29-30}: i\varepsilon_{\mu\nu\lambda\rho}\bar{T}^{abc,\eta}\langle f_+^{\mu\nu}\rangle  u^{ae,\lambda}D^\rho T_\eta^{bce}$  &&	\\
	11 & $i \varepsilon_{\mu\nu\lambda\rho}\psib\la \tilde{f}_+^{\mu\nu}u^{\lambda}\ra D^{\rho}\psi$&17 & $e^{(3)}_{24+26-25}:  i\varepsilon_{\mu\nu\lambda\rho}\bar{T}^{abc,\eta}\langle\tilde{f}_+^{\mu\nu}u^\lambda\rangle D^\rho T_\eta^{abc}$ & &\\
		&& 18 &$e_{25}^{(3)}: i\bar{T}^{abc}_\mu\langle \tilde{f}_+^{\mu\nu}u^\lambda\rangle\gamma_\lambda\gamma_5 T^{abc}_\nu$ &12&$f^{(3)}_{22}:  [i\epsilon^{  b c}\Nb^d\tilde{f}_+^{be,\mu\nu}u^{df}_\mu T^{cef}_\nu]_+$ \\
	&&19 &$e_{26-24}^{(3)}: [i\bar{T}^{abc}_\lambda[\tilde{f}^{\mu\nu}_+,u^\lambda]^{ae}\gamma_\mu\gamma_5 T^{bce}_\nu]_+$&13&$f^{(3)}_{27}: [i\epsilon^{  b c}\Nb^d\langle f_+^{\mu\nu}\rangle u^{be}_{\mu}T^{cde}_\nu]_+$\\
	& &20 &$e^{(3)}_{22-23}: i\varepsilon_{\mu\nu\lambda\rho}\bar{T}^{abc,\eta}\tilde{f}_+^{ad,\mu\nu}u^{be,\lambda}D^\rho T_\eta^{cde}$&14&$f^{(3)}_{29}: [\epsilon^{  b c}\Nb^d\langle f_+^{\mu\nu}\rangle u^{be,\lambda}\sigma_{\mu\nu}T^{cde}_{\lambda}]_+$\\
	&&21&$e^{(3)}_{23}: i\bar{T}^{abc}_\mu\tilde{f}_+^{ad,\mu\nu}u^{be,\lambda}\gamma_\lambda\gamma_5 T^{cde}_\nu$&15&$f^{(3)}_{26}: [\epsilon^{bc}\Nb^d\tilde{f}_+^{be,\mu\nu}u^{df,\lambda}\sigma_{\mu\nu}T^{cef}_\lambda]_+$\\
	 &&22 &$e^{(3)}_{30}: i\bar{T}^{abc}_\mu\langle f_+^{\mu\nu}\rangle u^{ae,\lambda}\gamma_\lambda\gamma_5 T^{bce}_\nu$ &16& $f^{(3)}_{25}: [\epsilon^{bc}\Nb^d[\tilde{f}_+^{\mu\nu},u^\lambda]^{be}\sigma_{\mu\nu}T^{cde}_\lambda]_+$\\
	\hline
	12 & $i \psib\nabla_{\mu}\tilde{f}_+^{\mu\nu}D_{\nu}\psi$& 23& $e^{(3)}_{31}: i\Tb^{ a b c,\rho}(\nabla_{\mu}\tilde{f}_+^{\mu\nu})^{ae}D_\nu T^{bce}_\rho$&&\\
	13 & $i \psib\la\nabla_{\mu}f_+^{\mu\nu}\ra D_{\nu}\psi$&24 &$e^{(3)}_{32}: i\Tb^{ a b c,\rho}\langle \nabla_\mu f_+^{\mu\nu}\rangle D_\nu T^{abc}_\rho$ &&  \\\hline
	14 & $[i \psib \tilde{f}_{+}^{\mu\nu}u^{\lambda} \gamma_{\mu}\gamf D_{\nu\lambda}\psi]_+$ & 25 & $e^{(3)}_{28}: [i\Tb^{ a b c,\rho}(\tilde{f}_+^{\mu\nu}u^\lambda)^{ae}\gamma_\mu\gamf D_{\nu\lambda}T^{bce}_\rho]_+$&17&$f^{(3)}_{23}: [i\epsilon^{  b c}\Nb^d[\tilde{f}_+^{\mu\nu},u^\lambda]^{be}D_{\nu\lambda}T^{cde}_\mu]_+$\\
	&&&& 18&$f^{(3)}_{24}: [i\epsilon^{  b c}\Nb^d\tilde{f}_+^{be,\mu\nu}u^{df,\lambda}D_{\nu\lambda}T^{cef}_\mu]_+$\\
&&&&19&$f^{(3)}_{28}: [i\epsilon^{  b c}\Nb^d\langle f_+^{\mu\nu}\rangle u^{be,\lambda}D_{\nu\lambda}T^{cde}_\mu]_+$\\
\hline
	15 & $[\psib u_{\mu}f_{-}^{\mu\nu}D_{\nu}\psi]_+$ & 26 &$e_{17}^{(3)}: [\Tb^{ a b c,\rho}(u_\mu f_-^{\mu\nu})^{ae}D_{\nu}T^{bce}_\rho]_+$ &&\\
	16 & $i \psib\la u^{\mu}f_{-}^{\nu\lambda}\ra\sigma_{\mu\nu}D_{\lambda}\psi$ & 27&$e^{(3)}_{14+16}: i\bar{T}^{abc,\rho}\langle u^\mu f_-^{\nu\lambda}\rangle\sigma_{\mu\nu}D_\lambda T^{abc}_\rho$&&\\
	17 & $i \psib\la u^{\mu}f_{-}^{\nu\lambda}\ra\sigma_{\nu\lambda}D_{\mu}\psi$ & 28&$e_{15}^{(3)}: i\Tb^{ a b c,\rho} \langle u^\mu f_-^{\nu\lambda}\rangle\sigma_{\nu\lambda} D_{\mu}T^{abc}_\rho$&&\\
	&&29	&$e_{14-16}^{(3)}: [\bar{T}^{abc}_\mu[u^\mu,f_-^{\nu\lambda}]^{ae}D_\lambda T^{bce}_\nu]_+$&20&	$f^{(3)}_{9+11}: [\epsilon^{bc}\Nb^d (u^\mu f_-^{\nu\lambda})^{be}D_\lambda(\gamma_{\mu}\gamf T^{cde}_\nu-\gamma_\nu\gamf T^{cde}_\mu)]_+$\\
&&30& $e_{12}^{(3)}: i\Tb^{ a b c,\rho}u^{ad,\mu} f_-^{be, \nu\lambda}\sigma_{\mu\nu}D_{\lambda}T^{ c d e}_\rho$  &  21&$f^{(3)}_{10}: [\epsilon^{  b c}\Nb^d (u^\mu f_-^{\nu\lambda})^{be}D_\mu(\gamma_{\nu}\gamf T^{cde}_\lambda-\gamma_\lambda\gamf T^{cde}_\nu)]_+$ \\
&&31&$e_{13}^{(3)}: i\Tb^{ a b c,\rho} u^{ad,\mu} f_-^{be,\nu\lambda}\sigma_{\nu\lambda}D_{\mu}T^{cde}_\rho $  & 22&$f^{(3)}_{9-11}: [\epsilon^{bc}\Nb^d (u^\mu f_-^{\nu\lambda})^{be}D_\lambda(\gamma_{\mu}\gamf T^{cde}_\nu+\gamma_\nu\gamf T^{cde}_\mu)]_+$\\
&&&   & 23&$f^{(3)}_{12+14}: [\epsilon^{ b c}\Nb^d u^{be,\mu}f_-^{df,\nu\lambda}D_\lambda(\gamma_\mu\gamf T^{cef}_\nu-\gamma_\nu\gamf T^{cef}_\mu)]_+$\\
&&&   & 24&$f^{(3)}_{13}: [\epsilon^{  b c}\Nb^du^{be,\mu}f_-^{df,\nu\lambda}D_\mu(\gamma_\nu\gamf T^{cef}_\lambda-\gamma_\lambda\gamf T^{cef}_\nu)]_+$\\
&&&   & 25&$f^{(3)}_{12-14}: [\epsilon^{ b c}\Nb^d u^{be,\mu}f_-^{df,\nu\lambda}D_\lambda(\gamma_\mu\gamf T^{cef}_\nu+\gamma_\nu\gamf T^{cef}_\mu)]_+$\\
&&&  &26&$f^{(3)}_{17}: [\varepsilon_{\mu\nu\lambda\rho}\epsilon^{bc}\Nb^d(u^\mu f_-^{\nu\lambda})^{be}T^{cde,\rho}]_+$\\
&&&  &27&$f^{(3)}_{18}: [\varepsilon_{\mu\nu\lambda\rho}\epsilon^{  b c}\Nb^du^{be,\mu} f_-^{df,\nu\lambda}T^{ c e f,\rho}]_+$\\
\hline
	18 & $\psib\nabla_{\mu}f_{-}^{\mu\nu} \gamma_{\nu}\gamf\psi$ & 32 & $e^{(3)}_{21}: \Tb^{ a b c,\rho}(\nabla_\mu f_-^{\mu\nu})^{ae}\gamma_{\nu}\gamf T^{bce}_\rho$& 28&$f^{(3)}_{19}: [\epsilon^{  b c}\Nb^d(\nabla_\mu f_-^{\mu\nu})^{be}T^{cde}_\nu]_+$\\
&&&& 29 &$f^{(3)}_{20}: [i\epsilon^{  b c}\Nb^d(\nabla^{\mu}f_-^{\nu\lambda})^{be}\sigma_{\mu\nu}T^{cde}_\lambda]_+$\\
\hline
	19 & $\psib\la u^{\mu}\tilde{\chi}_+\ra \gamma_{\mu}\gamf\psi$ &33 &$e^{(3)}_{34}: \Tb^{ a b c,\rho}\langle u^\mu\tilde{\chi}_+\rangle\gamma_{\mu}\gamf T^{abc}_\rho$ &&\\	
	20 & $\psib\la\chip\ra u^{\mu} \gamma_{\mu}\gamf\psi$ &34 &$e^{(3)}_{35}: \Tb^{ a b c,\rho}\langle\chi_+\rangle u^{ae,\mu}\gamma_{\mu}\gamf T^{abc}_\rho$   &30&$f^{(3)}_{32}: [\epsilon^{  b c}\Nb^d\langle \chi_+\rangle u^{be,\mu}T^{cde}_\mu]_+$\\
&&35 &$e^{(3)}_{33}: \Tb^{ a b c,\rho}u^{ad,\mu}\tilde{\chi}_+^{be}\gamma_{\mu}\gamf T^{cde}_\rho$ &31&$f^{(3)}_{30}: [\epsilon^{bc}\Nb^d(u^\mu\tilde{\chi}_+)^{be}T^{cde}_\mu]_+$\\
&&     & &32&$f^{(3)}_{31}: [\epsilon^{  b c}\Nb^d u^{be,\mu}\tilde{\chi}_+^{df}T^{cef}_\mu]_+$\\
\hline
	21 & $i \psib\tilde{\chi}_-^{\mu} \gamma_{\mu}\gamf\psi$    &36 &$e^{(3)}_{37}: i\Tb^{ a b c\rho}(\tilde{\chi}_-^{\mu})^{ae}\gamma_{\mu}\gamf T^{bce}_\rho$ & 33&$f^{(3)}_{33}: [i\epsilon^{  b c}\Nb^d(\tilde{\chi}_-^\mu)^{be}T^{cde}_\mu]_+$\\
	22 & $i \psib\la\chi_{-}^{\mu}\ra \gamma_{\mu}\gamf\psi$ &37 &$e^{(3)}_{38}: i\Tb^{ a b c,\rho}\langle\chi_-^\mu\rangle\gamma_{\mu}\gamf T^{abc}_\rho$ &&\\\hline
	23 & $[i \psib u^{\mu}\tilde{\chi}_- D_{\mu}\psi]_+$ &38&$e^{(3)}_{36}: [i\Tb^{ a b c,\rho}(u^\mu\tilde{\chi}_-)^{ae}D_{\mu}T^{bce}_\rho]_+$ &&\\
	\hline\hline
\end{tabular}
\end{table*}

In Tables \ref{p1}-\ref{p3}, the terms of ${\cal L}_{\pi NN/\pi qq}$ and the names of the coupling coefficients are the same as those in Ref. \cite{Jiang:2022gjy}. The $\Delta$ interaction terms are reorganized from those in Ref. \cite{Jiang:2018mzd} so that we may set structure correspondences between the hadron- and quark-level descriptions and get the LEC relations easily. In the leading order $\pi\Delta\Delta$ ($\pi N\Delta$) Lagrangian, we use $e_1^{(1)}$ ($f_1^{(1)}$) to denote  the coupling constant. Since the ${\cal O}(p^2)$ and ${\cal O}(p^3)$ terms in  $\pi\Delta\Delta$ ($\pi N\Delta$) Lagrangians are reorganized, the related coupling constants  are called $\hat{e}_i^{(2,3)}$ ($\hat{f}_i^{(2,3)}$) now. Note that all the $f$ coefficients are real. To see the relations between the reorganized terms and the original ones, we also show additional symbols in front of each term.  The symbol $f^{(2)}_{1-2-2}$ means that the $\hat{f}_1^{(2)}$ term is from the following combination of the $f^{(2)}_1$ and $f^{(2)}_2$ terms constructed in Ref. \cite{Jiang:2018mzd}: [($f^{(2)}_1$ term)$-$($f^{(2)}_2$ term)$-$($f^{(2)}_2$ term)]. One understands other symbols similarly. However, we just replace the original 18th (15th) term of ${\cal O}(p^3)$ ${\cal L}_{\pi\Delta\Delta}$ with the new 14th (9th) term so that we may discuss the LEC relations with the structure correspondences shown in Sec. \ref{sec4}.

When reorginazing the $\Delta$ interaction terms, we have adopted
\begin{eqnarray}
&&[X,Y]^{af}Z^{be}+[Y,Z]^{af}Y^{be}+[Z,X]^{af}Y^{be}\nonumber\\
&=&\langle XYZ\rangle(2\delta^{ae}\delta^{bf}-\delta^{af}\delta^{be})
\end{eqnarray}
where $X$, $Y$, and $Z$ are traceless 2 by 2 matrices, and
\begin{eqnarray}
&&\epsilon^{ac}(X^{ab}Y^{cd}-Y^{ab}X^{cd})\nonumber\\
&=&\epsilon^{ac}\Big[\langle X\rangle \delta^{ab}Y^{cd}+\langle Y\rangle X^{ab}\delta^{cd}+(XY-YX)^{ab}\delta^{cd}\Big]
\end{eqnarray}
where $X$ and $Y$ are arbitrary 2 by 2 matrices and $b$ and $d$ are symmetric indices. Several properties in the following Eqs. (\ref{St1})-(\ref{St6}) are also used.

\section{Quark wave functions of baryons and transition spin for $\Delta$ field}\label{sec3}

\begin{figure}[htbp]
\centering
\includegraphics{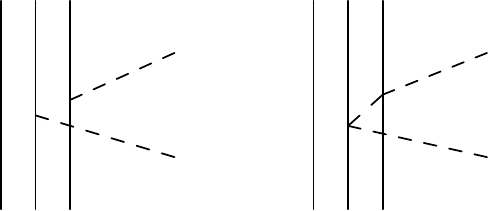}
\caption{Possible corrections to a $\Delta$-$\Delta$-boson or $N$-$\Delta$-boson coupling at the quark level.}\label{appcor}
\end{figure}

In the same spirit of Ref. \cite{Jiang:2022gjy}, one needs to find LEC relations by comparing hadron- and quark-level coupling matrix elements in the nonrelativistic limit. When we say {\it boson} for a coupling in the following discussions, it means pion or building block $u^\mu$, $\chi_\pm$, $\chi^\mu_\pm$, $h^{\mu\nu}$, or $f^{\mu\nu}_\pm$. The nonvanishing structures of $B$-$B^\prime$-boson couplings in the nonrelativistic limit have the reduction forms
\begin{eqnarray}
\bar{B}B^\prime&\to& B_H^\dag B^\prime_H,\nonumber\\
\bar{B}\gamma^\mu\gamma_5B^\prime&\to& B_H^\dag\sigma^k B^\prime_H\quad (\mu=k=1,2,3), \nonumber\\
\bar{B}\sigma^{\mu\nu}B^\prime&\to& \epsilon^{ijk}B_H^\dag\sigma^k B^\prime_H\quad (\mu=i,\nu=j),
\end{eqnarray}
where $B$ and $B^\prime$ can be nucleon, $\Delta$, or quark. We still assume that a baryon-baryon-boson coupling at the hadron level is described by a quark-quark-boson coupling at the quark level and two of the three quarks in $N$ or $\Delta$ are just spectators. Beyond this assumption, the corrections from the one-quark spectator case and the case without spectator quark should also contribute to the final LEC relations. When there is only one external boson (leading order term), only the two-quark spectator case is involved for the baryon-baryon-boson coupling. The coupling constant $g_A^q=\frac35g_A$ in the usual chiral quark models \cite{Manohar:1983md,Zhang:1994pp,Zhang:1997ny} is determined in this way. The correctness of the chiral quark model can be verified through the successful studies of baryon-baryon scattering processes. When there are two external bosons for the baryon-baryon-boson coupling (${\cal O}(p^2)$ terms), the one-quark spectator case may be involved where the two bosons are connected to different quark lines. Fig. \ref{appcor} illustrates two possible correction diagrams at the quark level. When there are three external bosons (${\cal O}(p^3)$ terms), the corrections from both the case with one-quark spectator and the case without spectator quark (the three bosons are connected to different quark lines) may be involved. As a first step for the LEC relation study, here we consider only the two-quark spectator case. In fact, the final LEC relations may also get contributions from chiral corrections. From the numerical analysis in our previous study \cite{Jiang:2022gjy}, it seems that the importance of the corrections grows for high order terms. We will defer the consideration of the mentioned corrections to future studies.

In Ref. \cite{Jiang:2022gjy}, we explicitly wrote down the quark-level spin-flavor wave functions of the octet baryons in order to confirm the LEC relations with a computer program. In the present study, the nucleon wave functions are still needed. At the quark level, they can be simply denoted as
\begin{eqnarray}\label{NWF}
N^i_\alpha = \sum_{xyz;\rho\xi\eta} W^{i,xyz}_{\alpha,\rho\xi\eta}q^x_\rho q^y_\xi q^z_\eta
\end{eqnarray}
where $i,x,y,z$ are flavor indices (1 represents $p$ or $u$ and 2 represents $n$ or $d$) and $\alpha,\rho,\xi,\eta$ are spin indices (1 represents $\uparrow$ and 2 represents $\downarrow$). For the three quarks in $\Delta$, the spin of any two quarks is 1 and it is easy to write down the quark spin wave functions explicitly. At the hadron level, to express the coupling matrix elements in an operator form, we adopt the notation of transition spin $S_t^\mu$. It is defined through
\begin{eqnarray}
&&T_\mu^{abc}\equiv S_{t\mu}T^{abc}=\sqrt{\frac13(\frac92-a-b-c)^2+\frac14}\, \Phi_\mu^{abc}\nonumber\\
&=&\sqrt{\frac13(\frac92-a-b-c)^2+\frac14}\, S_{t\mu}\Phi^{abc}.
\end{eqnarray}
The factor $\sqrt{\frac13(\frac92-a-b-c)^2+\frac14}$ just represents the number 1 or $\frac{1}{\sqrt3}$ in (\ref{T111}). In the static limit, $S_t^\mu=(0,\vec{S}_t)$ and the third components for the $\Delta$ spin wave function read
\begin{eqnarray}
&\Phi_{s_z=3/2}=\left(\begin{array}{c}1\\0\\0\\0\end{array}\right),\quad
\Phi_{s_z=1/2}=\left(\begin{array}{c}0\\1\\0\\0\end{array}\right),&\nonumber\\
&\Phi_{s_z=-1/2}=\left(\begin{array}{c}0\\0\\1\\0\end{array}\right),\quad
\Phi_{s_z=-3/2}=\left(\begin{array}{c}0\\0\\0\\1\end{array}\right).&
\end{eqnarray}
Using the RS spinor expression $\Phi_\mu(s_z)=\sum_{\lambda,s}\langle 1\lambda\frac12 s|\frac32 s_z\rangle$
$\times e_\mu(\lambda)u(s)$ \cite{Hemmert:1997ye}, one gets
\begin{eqnarray}\label{St}
S_t^1&=&\left(\begin{array}{cccc}-\frac{1}{\sqrt2}&0&\frac{1}{\sqrt6}&0\\0&-\frac{1}{\sqrt6}&0&\frac{1}{\sqrt2}\end{array}\right),\nonumber\\
S_t^2&=&\left(\begin{array}{cccc}-\frac{i}{\sqrt2}&0&-\frac{i}{\sqrt6}&0\\0&-\frac{i}{\sqrt6}&0&-\frac{i}{\sqrt2}\end{array}\right), \nonumber\\
S_t^3&=&\left(\begin{array}{cccc}0&\sqrt{\frac{2}{3}}&0&0\\0&0&\sqrt{\frac{2}{3}}&0\end{array}\right).
\end{eqnarray}
At the quark level, similar to Eq. (\ref{NWF}), a $\Delta$ state can be denoted as
\begin{eqnarray}\label{Deltaw.f.}
\Phi^{abc}_\alpha&=&\sum_{x,y,z;\rho,\tau,\eta} V^{abc,xyz}_{\alpha,\rho\tau\eta}q^x_\rho q^y_\tau q^z_\eta\nonumber
\\
&=&\sqrt{\frac13(\frac52-\alpha)^2+\frac14}\sum_{x,y,z;\rho,\tau,\eta} \delta_{a+b+c,x+y+z}\delta_{\alpha+2,\rho+\tau+\eta}\nonumber\\
&&\times  q^x_\rho q^y_\tau q^z_\eta,
\end{eqnarray}
where $\alpha=1,2,3,4$ represent $s_z=3/2,1/2,-1/2,-3/2$, respectively.

We will use the following properties for the transition spin (correct up to a higher chiral order):
\begin{eqnarray}
S_{t\beta}&\doteq&i\sigma_{\beta\nu}S_t^\nu,\label{St1}\\
\gamma^\mu\gamma^5S_t^\nu-\gamma^\nu\gamma^5S_t^\mu&\doteq&i\varepsilon^{\mu\nu\lambda\rho}\gamma_\rho S_{t\lambda}\nonumber\\
&\doteq& -\frac{1}{m_\Delta}\varepsilon^{\mu\nu\lambda\rho}D_\rho S_{t\lambda},\label{St2}\\
\sigma^{\mu\nu}T^\lambda&\doteq&\sigma^{\mu\lambda}T^\nu+\sigma^{\lambda\nu}T^\mu,\label{St3}\\
\bar{S}_{t\mu}S^\mu_t&=&-1,\label{St4}\\
i\varepsilon_{\mu\nu\lambda\rho}\bar{S}_t^\eta\gamma^\lambda S_{t\eta}
&\doteq&\bar{S}_{t\mu}\gamma_\rho\gamma^5 S_{t\nu}+\bar{S}_{t\nu}\gamma_\mu\gamma^5 S_{t\rho}\nonumber\\
&&+\bar{S}_{t\rho}\gamma_\nu\gamma^5 S_{t\mu}-\bar{S}_{t\mu}\gamma_\nu\gamma^5 S_{t\rho}\nonumber\\
&&-\bar{S}_{t\nu}\gamma_\rho\gamma^5 S_{t\mu}-\bar{S}_{t\rho}\gamma_\mu\gamma^5 S_{t\nu}\nonumber\\
&\doteq& -\frac{1}{m_\Delta}\varepsilon_{\mu\nu\lambda\rho}\bar{S}_t^\eta D^\lambda S_{t\eta},\label{St5}\\
i(\bar{S}_t^\mu S_t^\nu-\bar{S}_t^\nu S_t^\mu)&\doteq&\bar{S}_t^\rho\sigma^{\mu\nu}S_{t,\rho}.  \label{St6}
\end{eqnarray}
where $\bar{S}_{t\mu}\equiv S_{t\mu}^\dag\gamma^0$. In getting these equations, the subsidiary condition $\gamma_\mu S_t^\mu\doteq 0$ has been employed.

\section{Low-energy constant relations}\label{sec4}

Since the quark flavors are explicitly shown in the $\Delta$ field, one may only focus on the Lorentz structure correspondences. From Tables \ref{p1}-\ref{p3}, it is obvious that the one-to-one correspondence between the hadron-level interaction terms and the quark-level terms is not always there. The number of independent terms in ${\cal L}_{\pi\Delta\Delta}$ is more than that in ${\cal L}_{\pi qq}$ while the number of independent terms in ${\cal O}(p^2)$ (${\cal O}(p^3)$) ${\cal L}_{\pi N\Delta}$ is less (more) than that in ${\cal L}_{\pi qq}$. From the ${\cal O}(p^1)$ and ${\cal O}(p^2)$ order Lagrangian terms, one sets the following five structure correspondences:
\begin{eqnarray}
k_1S_t^\mu&\to&\gamma^\mu\gamma^5,\label{cor1}
\\
k_2\sigma_t^{\mu\nu}\equiv ik_2(\gamma^\mu\gamma^5S_t^\nu-\gamma^\nu\gamma^5S_t^\mu)&&\nonumber\\
=k_2\epsilon^{\mu\nu\rho\lambda}\gamma_\rho S_{t\lambda}&\to&\sigma^{\mu\nu};\label{cor2}
\\
-\bar{S}_{t\rho}S_t^\rho=1&\to& 1,\label{cor3}
\\
\sigma_{RS}^\mu\equiv-\bar{S}_{t\rho}\gamma^\mu\gamma^5 S_t^\rho&\to & \gamma^\mu\gamma^5,\label{cor4}
\\
\sigma_{RS}^{\mu\nu}\equiv -i(\bar{S}_t^\mu S_t^\nu-\bar{S}_t^\nu S_t^\mu)=-\bar{S}_{t\rho}\sigma^{\mu\nu}S_t^\rho& \to&\sigma^{\mu\nu},\label{cor5}
\end{eqnarray}
where $k_1$ and $k_2$ are constants to be discussed in next section. The first two (latter three) will be used to get LEC relations between ${\cal L}_{\pi N\Delta}$ and ${\cal L}_{\pi qq}$ (${\cal L}_{\pi \Delta\Delta}$ and ${\cal L}_{\pi qq}$). At the third chiral order, additional structures appear. We set
\begin{eqnarray}
k_3\sigma^{\mu\nu}S_t^\lambda
&\to& i\Big(g^{\nu\lambda}\gamma^\mu\gamma^5-g^{\mu\lambda}\gamma^\nu\gamma^5\Big)\nonumber\\
&&+\varepsilon^{\mu\nu\lambda\rho}\gamma_\rho\nonumber\\
&\doteq& i\Big(g^{\nu\lambda}\gamma^\mu\gamma^5-g^{\mu\lambda}\gamma^\nu\gamma^5\Big)\nonumber\\
&&+\frac{i}{m_\Delta}\varepsilon^{\mu\nu\lambda\rho}D_\rho,\label{cor6}
\\
k_4(\gamma^\mu\gamma_5S_t^\nu+\gamma^\nu\gamma_5S_t^\mu)&\to&g^{\mu\nu};\label{cor7}
\\
-k_5(\bar{S}_t^\mu S_t^\nu+\bar{S}_t^\nu S_t^\mu)&\to&g^{\mu\nu},\label{cor8}
\\
-k_6(\bar{S}_t^\mu \gamma^\lambda\gamma_5S_t^\nu+\bar{S}_t^\nu \gamma^\lambda\gamma_5S_t^\mu)&\to&g^{\lambda\mu}\gamma^\nu\gamma_5+g^{\lambda\nu}\gamma^\mu\gamma_5\nonumber\\
&&-g^{\mu\nu}\gamma^\lambda\gamma^5,\label{cor9}
\\
-k_7(\bar{S}_t^\mu \gamma^\lambda\gamma_5S_t^\nu-\bar{S}_t^\nu \gamma^\lambda\gamma_5S_t^\mu)&\to&i\varepsilon^{\mu\nu\lambda\rho}\gamma_\rho\nonumber\\
&\doteq&-\frac{1}{m_\Delta}\varepsilon^{\mu\nu\lambda\rho}D_\rho.\label{cor10}
\end{eqnarray}
Again, the first two (latter three) will be used to get LEC relations between ${\cal L}_{\pi N\Delta}$ and ${\cal L}_{\pi qq}$ (${\cal L}_{\pi \Delta\Delta}$ and ${\cal L}_{\pi qq}$). They are obtained by employing the correspondence (\ref{cor1}). In this procedure, the multiplication factor may be changed and we adopt new constants $k_3$, $\cdots$, and $k_7$. The following LEC relations to be discussed are based on these correspondences.

Before further discussions, let us take a look at the vanishing hadron-level structures according to the structure correspondences. In Tables \ref{p1}-\ref{p3}, only those terms that can be directly connected with the above correspondences (\ref{cor1})-(\ref{cor5}) are shown in the same row. Obviously, each $\pi qq$ term has a corresponding $\Delta$-$\Delta$-boson counterpart, but it is not the case for the $N$-$\Delta$-boson couplings.
Scrutinizing the structures of the ${\cal O}(p^3)$ $\pi qq$ terms without $N$-$\Delta$-boson counterparts, one sees that they can be classified into two types: 1) terms without gamma matrices but with at least one derivative $D^\mu$ and 2) terms that all building blocks $u^\mu$, $h^{\mu\nu}$, $f_\pm^{\mu\nu}$, $\chi_\pm$, and $\chi_-^\mu$ are placed in the same trace. In the former case, the derivative can be approximately replaced with $\gamma^\mu=-\gamma^5(\gamma^\mu\gamma^5)$. Then the $\pi N\Delta$ term according to the correspondence (\ref{cor1}) should have a higher order structure with $\gamma^5$ and there is no ${\cal O}(p^3)$ $\pi N\Delta$ counterpart. In the later case, the corresponding $N$-$\Delta$-boson terms are equal to zero because of the existence of both antisymmetric tensor $\epsilon^{bc}$ and symmetric tensor $T^{bcd}$, e.g. the counterpart of the $\beta_8^{(3)}$ term reads $[\epsilon^{bc}\bar{N}^d\langle u^\mu h^{\nu\lambda}\rangle D_\lambda(\gamma_\mu\gamma^5 T^{bcd}_\nu-\gamma_\nu\gamma^5 T^{bcd}_\mu)]_+=0$. For the ${\cal O}(p^2)$ case, the situation is slightly different. A trace always exists in the $\beta_1^{(2)}$, $\beta_3^{(2)}$, $\beta_5^{(2)}$, and $\beta_7^{(2)}$ terms and there are no structure counterparts in ${\cal L}_{\pi N\Delta}$. A nonvanishing $N$-$\Delta$-boson term corresponding to the $\beta_6^{(2)}$ term should look like $[\epsilon^{ab}\bar{N}^c(\tilde{\chi}_+)^{ae}\Gamma^\mu T^{bce}_\mu]_+$ with $\Gamma^\mu$ being a combination of $\gamma$ matrices. Such a structure, however, does not exist because of the subsidiary condition $\gamma^\mu T_\mu\doteq0$.

\subsection{${\cal O}(p^1)$ relations}

It is easy to guess that $e_1^{(1)}=-\frac32g_A^q$ since $\bar{S}_{t\rho}S_t^\rho=-1$ and there are three possibilities for the external pion couples with a quark inside $\Delta$. As an example for the calculation with structure correspondence, we focus on the $\Delta^{++}_{s_z=3/2}$-$\Delta^{++}_{s_z=3/2}$-$\pi^0$ coupling vertex. The explicit expression for the matrix element at the hadron level is
\begin{eqnarray}
{\cal M}\sim -e_1^{(1)} \langle \Delta^{++}_{s_z=3/2}|\sigma_{RS,3}(\tau_3)^{11}q_3|\Delta^{++}_{s_z=3/2}\rangle =-e_1^{(1)} q_3,\nonumber\\
\end{eqnarray}
where $q_3$ is the $z$-component of pion momentum, $\tau_3$ is the third Pauli matrix, and $\sigma_{RS,3}$ is the third component of the nonrelativistic form of $\sigma_{RS}^\mu$ defined in the correspondence (\ref{cor4}). The same vertex at the quark level can be calculated with Eq. (\ref{Deltaw.f.})
\begin{eqnarray}
{\cal M}\sim \frac12g_A^q\langle \Delta^{++}_{s_z=3/2}|\sum_{i=1}^3\sigma^{(i)}_3\tau^{(i)}_3q_3|\Delta^{++}_{s_z=3/2}\rangle=\frac32g_A^q q_3,\nonumber\\
\end{eqnarray}
where $i$ labels quarks and $\sigma_3$ is also the third Pauli matrix. Therefore, one has $e_1^{(1)}=-\frac32g_A^q$ from the correspondence (\ref{cor4}). This relation can be obtained with each coupling term and it is consistent with that given in the literature.

Similarly, one finds $f_1^{(1)}=k_1g_A^q$ with the correspondence (\ref{cor1}). Although the real $f_1^{(1)}$ can be positive or negative, we here choose its sign to be consistent with this relation\footnote{In the following discussions about $\hat{f}_i^{(2)}$ and $\hat{f}_i^{(3)}$, their signs are similarly chosen.}. To confirm this relation with a program, we consider the coupling vertex of $\Phi^{bcd}_\alpha\to N^c_\beta+\pi^n$. Its matrix element at the hadron level reads
\begin{eqnarray}
&&{\cal M}(b+c+d,\alpha\to c,\beta; p^m,\pi^n)\nonumber\\
&=&-f_1^{(1)}\sqrt{\frac13(\frac92-b-c-d)^2+\frac14}\nonumber\\
&&\times\sum_{b\leftrightarrow d}\sum_a\Big[\epsilon^{ab}\langle N^c_\beta| S_t^m (\tau^n)^{ad}|\Phi^{bcd}_\alpha\rangle\Big]\quad\\
&=&-f_1^{(1)}\sqrt{\frac13(\frac92-b-c-d)^2+\frac14}\sum_{b\leftrightarrow d}\Big[(S_t^m)_{\beta\alpha} (-i\tau^2\tau^n)^{bd}\Big],\nonumber\\
\end{eqnarray}
where $S_t^m$ is given in Eq. (\ref{St}). The quark-level description is
\begin{eqnarray}
&&{\cal M}(b+c+d,\alpha\to c,\beta; p^m,\pi^n)\nonumber\\
&=& -3f_1^q\sum^{x'xyz}_{\rho'\rho\tau\eta} W^{c,x'yz}_{\beta,\rho'\tau\eta}V^{bcd,xyz}_{\alpha,\rho\tau\eta}\langle{q^{x'}_{\rho'}}|(\sigma^m)(\tau^n) |q^x_\rho\rangle\\
&=& -\frac32g_A^q\sum^{x'xyz}_{\rho'\rho\tau\eta} W^{c,x'yz}_{\beta,\rho'\tau\eta}\delta_{b+c+d,x+y+z}\delta_{2+\alpha,\rho+\tau+\eta}\nonumber\\
&&\times\sqrt{\frac13(\frac92-b-c-d)^2+\frac14}\nonumber\\
&&\times\sqrt{\frac13(\frac52-\alpha)^2+\frac14} (\sigma^m)_{\rho'\rho}(\tau^n)^{x'x}.
\end{eqnarray}
From the equivalence of these two expressions for various vertices, or from the vanishing of the function
\begin{eqnarray}
&&G(b+c+d,\alpha\to c,\beta; p^m,\pi^n)\nonumber\\
&\equiv&f_1^{(1)}(S_t^m)_{\beta\alpha}\sum_{b\leftrightarrow d}(-i\tau^2\tau^n)^{bd}
-\frac32k_1g_A^q\sqrt{\frac13(\frac52-\alpha)^2+\frac14}\nonumber\\
&&\times\sum^{x'xyz}_{\rho'\rho\xi\eta} W^{c,x'yz}_{\beta,\rho'\xi\eta}\delta_{b+c+d,x+y+z}\delta_{2+\alpha,\rho+\xi+\eta} (\sigma^m)_{\rho'\rho}(\tau^n)^{x'x},\nonumber\\
\end{eqnarray}
 one can confirm $f_1^{(1)}=k_1g_A^q$.

\subsection{${\cal O}(p^2)$ relations}

At this order, according to the correspondences (\ref{cor3}) and (\ref{cor5}) and the nonrelativistic form of $\sigma_{RS}^{\mu\nu}$, one may get
\begin{eqnarray}
&\hat{e}_1^{(2)}=-3\beta_1^{(2)},\, \hat{e}_5^{(2)}=-3\beta_2^{(2)},\, \hat{e}_6^{(2)}=-3\beta_3^{(2)},\, \hat{e}_8^{(2)}=-3\beta_4^{(2)},&\nonumber\\
&\hat{e}_9^{(2)}=-3\beta_5^{(2)},\, \hat{e}_{10}^{(2)}=-3\beta_6^{(2)},\, \hat{e}_{11}^{(2)}=-3\beta_7^{(2)}.&
\end{eqnarray}
For the $\hat{e}_2^{(2)}$, $\hat{e}_4^{(2)}$, and $\hat{e}_7^{(2)}$ terms, we set the coupling constants to be zero in the present study since they involve the case that two pions are coupled to different quark lines. We leave the relevant discussions in a future work. The determination of $\hat{e}_3^{(2)}$ depends on the correspondence (\ref{cor8}). It leads to $\hat{e}_3^{(2)}=-6k_5\beta_1^{(2)}$.

Similar to the procedure to determine $f_1^{(1)}$, the correspondence (\ref{cor2}) leads to the relations
\begin{eqnarray}
\hat{f}_2^{(2)}=2k_2\beta_2^{(2)},\qquad \hat{f}_3^{(2)}=2k_2\beta_4^{(2)}.
\end{eqnarray}

\subsection{${\cal O}(p^3)$ relations}

With the above experiences, one may classify the ${\cal O}(p^3)$ terms into three categories: 1) LECs can be obtained with the correspondences \eqref{cor1}-\eqref{cor5}, 2) LECs cannot be obtained with the assumption that two quarks inside the baryon are spectators, and 3) LECs can be obtained with the correspondences \eqref{cor6}-\eqref{cor10}. For terms in the second category, we set the hadron-level coefficients to be zero in this work. For terms in the first category, most relations between hadron-level coupling constants and quark-level coupling constants can be directly obtained, but the $\hat{f}_{13,19,20,21,26,31}^{(3)}$ terms are exceptions. For terms in the third category, one needs to explore them one by one. Here we discuss these exceptional terms.

In the $\Delta$-$\Delta$-boson case, all the LEC relations in the third category involve $k_5$, $k_6$, or $k_7$. The hadron-level $\hat{e}_6^{(3)}$ and $\hat{e}_7^{(3)}$ terms and quark-level $\beta_1^{(3)}$ and $\beta_2^{(3)}$ terms should be considered together since the same correspondence \eqref{cor9} is adopted. It results in $\hat{e}_6^{(3)}=-3k_6\beta_2^{(3)}$ and $\hat{e}_7^{(3)}=-3k_6(\beta_1^{(3)}+\frac12\beta_2^{(3)})$. Other LECs to be considered similarly are $\hat{e}_{18}^{(3)}$, $\hat{e}_{19}^{(3)}$, $\hat{e}_{22}^{(3)}$, and $\hat{e}_{29}^{(3)}$. Table \ref{LECrelations} lists the relevant results.

In the $N$-$\Delta$-boson case, the situation is slightly complicated. The relations $\hat{f}_5^{(3)}=k_3(\beta_1^{(3)}-\beta_2^{(3)})$, $\hat{f}_9^{(3)}=4k_4\beta_6^{(3)}$, and $\hat{f}_{22}^{(3)}=4k_4\beta_{15}^{(3)}$ can be trivially obtained with the correspondence \eqref{cor6} or \eqref{cor7}. To get the relation for $\hat{f}_{13}^{(3)}$, one may adopt the correspondence \eqref{cor1} directly,  which gives a non-Hermitian structure. It should be discarded. Alternatively, one may think that it can be eliminated by the structure from the Hermitian part of the hadron-level term. Then one has $\hat{f}_{13}^{(3)}=0$. Similarly, the results for $\hat{f}_{14,16,19,20,21,26,29,31}^{(3)}$ in Table \ref{LECrelations} are obtained.\\

\begin{table*}[htbp]\footnotesize
\caption{Obtained coupling constant relations between ${\cal L}_{\pi\Delta\Delta,\pi N\Delta}$ and ${\cal L}_{\pi qq}$.}\label{LECrelations}
\begin{tabular}{ccccc}\hline\hline
Chiral order&Group& ${\cal L}_{\pi\Delta\Delta}\Leftrightarrow {\cal L}_{\pi qq}$&${\cal L}_{\pi N\Delta}\Leftrightarrow {\cal L}_{\pi qq}$ \\\hline
  ${\cal O}(p^1)$ &1& $e_1^{(1)}=-\frac32g_A^q$. & $f_1^{(1)}=k_1g_A^q$.\\\hline
${\cal O}(p^2)$ &1& $\hat{e}_1^{(2)}=-3\beta_1^{(2)}$, $\hat{e}_{2,4}^{(2)}=0$, $\hat{e}_3^{(2)}=-6k_5\beta_1^{(2)}$;  & $\hat{f}_1^{(2)}=0$.\\
  &2&   $\hat{e}_5^{(2)}=-3\beta_2^{(2)}$; & $\hat{f}_2^{(2)}=2k_2\beta_2^{(2)}$\\
  &3&  $\hat{e}_6^{(2)}=-3\beta_3^{(2)}$, $\hat{e}_7^{(2)}=0$;  \\
  &4&   $\hat{e}_8^{(2)}=-3\beta_4^{(2)}$, $\hat{e}_9^{(2)}=-3\beta_5^{(2)}$; & $\hat{f}_3^{(2)}=2k_2\beta_4^{(2)}$.\\
  &5&   $\hat{e}_{10}^{(2)}=-3\beta_6^{(2)}$, $\hat{e}_{11}^{(2)}=-3\beta_7^{(2)}$. \\\hline
${\cal O}(p^3)$ &1& $\hat{e}_{1}^{(3)}=-3\beta_{1}^{(3)}$,  $\hat{e}_{2}^{(3)}=-3\beta_2^{(3)}$, $\hat{e}_{3}^{(3)}=-3\beta_3^{(3)}$, $\hat{e}_{4,5,8}^{(3)}=0$, & $\hat{f}_{1}^{(3)}=2k_1\beta_{1}^{(3)}$, $\hat{f}_{2}^{(3)}=2k_1\beta_{2}^{(3)}$, \\
&& $\hat{e}_{6}^{(3)}=-3k_6\beta_2^{(3)}$, $\hat{e}_{7}^{(3)}=-3k_6(\beta_1^{(3)}+\frac12\beta_2^{(3)})$;
&  $\hat{f}_{3,4}^{(3)}=0$, $\hat{f}_{5}^{(3)}=k_3(\beta_{1}^{(3)}-\beta_2^{(3)})$;\\
  &2&   $\hat{e}_{9}^{(3)}=-3\beta_4^{(3)}$, $\hat{e}_{10}^{(3)}=-3\beta_5^{(3)}$, $\hat{e}_{11}^{(3)}=0$; & $\hat{f}_{6}^{(3)}=2k_1\beta_{4}^{(3)}$, $\hat{f}_{7}^{(3)}=2k_1\beta_{5}^{(3)}$, $\hat{f}_{8}^{(3)}=0$;\\
  &3&   $\hat{e}_{12}^{(3)}=-3\beta_6^{(3)}$;\\
  &4& $\hat{e}_{13}^{(3)}=-3\beta_7^{(3)}$;\\
  &5& $\hat{e}_{14}^{(3)}=-3\beta_8^{(3)}$;  &  $\hat{f}_{9}^{(3)}=4k_4\beta_{6}^{(3)}$, $\hat{f}_{10}^{(3)}=0$;\\
  &6&   $\hat{e}_{15}^{(3)}=-3\beta_9^{(3)}$, $\hat{e}_{16}^{(3)}=-3\beta_{10}^{(3)}$, $\hat{e}_{17}^{(3)}=-3\beta_{11}^{(3)}$,  &  $\hat{f}_{11}^{(3)}=2k_1\beta_{9}^{(3)}$, $\hat{f}_{12,13,15}^{(3)}=0$, \\
  &&  $\hat{e}_{18}^{(3)}=6k_7m_\Delta\beta_{11}^{(3)}$, $\hat{e}_{19}^{(3)}=-\frac32k_6\beta_9^{(3)}$,   &  $\hat{f}_{14}^{(3)}=2k_3m_\Delta\beta_{10}^{(3)}$, $\hat{f}_{16}^{(3)}=-k_3\beta_{9}^{(3)}$;\\
  && $\hat{e}_{20,21}^{(3)}=0$, $\hat{e}_{22}^{(3)}=6k_7m_\Delta\beta_{10}^{(3)}$;\\
  &7&  $\hat{e}_{23}^{(3)}=-3\beta_{12}^{(3)}$, $\hat{e}_{24}^{(3)}=-3\beta_{13}^{(3)}$;\\
  &8&   $\hat{e}_{25}^{(3)}=-3\beta_{14}^{(3)}$;  & $\hat{f}_{17}^{(3)}=2k_1\beta_{14}^{(3)}$, $\hat{f}_{18,19}^{(3)}=0$;\\
  &9&   $\hat{e}_{26}^{(3)}=-3\beta_{15}^{(3)}$, $\hat{e}_{27}^{(3)}=-3\beta_{16}^{(3)}$,  & $\hat{f}_{20,21,23-27}^{(3)}=0$, $\hat{f}_{22}^{(3)}=4k_4\beta_{15}^{(3)}$; \\
  &&   $\hat{e}_{28}^{(3)}=-3\beta_{17}^{(3)}$, $\hat{e}_{29}^{(3)}=-3k_5\beta_{15}^{(3)}$, $\hat{e}_{30,31}^{(3)}=0$;  &  \\
  &10&   $\hat{e}_{32}^{(3)}=-3\beta_{18}^{(3)}$;  &  $\hat{f}_{28}^{(3)}=2k_1\beta_{18}^{(3)}$, $\hat{f}_{29}^{(3)}=2k_3\beta_{18}^{(3)}$;\\
  &11&   $\hat{e}_{33}^{(3)}=-3\beta_{19}^{(3)}$, $\hat{e}_{34}^{(3)}=-3\beta_{20}^{(3)}$, $\hat{e}_{35}^{(3)}=0$; &  $\hat{f}_{30}^{(3)}=2k_1\beta_{20}^{(3)}$, $\hat{f}_{31,32}^{(3)}=0$;\\
  &12&   $\hat{e}_{36}^{(3)}=-3\beta_{21}^{(3)}$, $\hat{e}_{37}^{(3)}=-3\beta_{22}^{(3)}$;   &  $\hat{f}_{33}^{(3)}=2k_1\beta_{21}^{(3)}$\\
  &13&   $\hat{e}_{38}^{(3)}=-3\beta_{23}^{(3)}$.\\
\hline\hline
\end{tabular}
\end{table*}

We summarize all the obtained LEC relations in Table \ref{LECrelations}. Both hadron-level and quark-level Lagrangians contain covariant derivative $D_\mu$ acting on matter fields. Since we treat the quark Lagrangian as an equivalent description of the hadron Lagrangians, $D_\mu$ acting on the quark field also leads to the scale $m_N$ or $m_\Delta$. Consequently, the numerical values of LECs in ${\cal L}_{\pi qq}$ may be different when ${\cal L}_{\pi qq}$ is treated as a quark-level description of ${\cal L}_{\pi NN}$, ${\cal L}_{\pi\Delta\Delta}$, or ${\cal L}_{\pi N\Delta}$. If the small scale expansion scheme is adopted in $\Delta\chi$PT, the differences are of high order. Then in this situation, the relations between LECs in ${\cal L}_{\pi NN}$, ${\cal L}_{\pi\Delta\Delta}$, and ${\cal L}_{\pi N\Delta}$ can be established by noticing the results shown in Table IV of Ref. \cite{Jiang:2022gjy}.

\section{Values of $k$'s}\label{sec5}

Since the spin operators for RS fields and for Dirac fields are different, we set several multiplication factors $k$'s in the structure correspondences \eqref{cor1}-\eqref{cor10}. One should adjust their values to ensure the equivalence of the hadron- and quark-level descriptions for the same coupling in the comparison. Now we move on to discuss these correspondences and relevant values of $k$'s.

The factor $k_1$ first appears in the leading order relation $f_1^{(1)}=k_1g_A^q$. Noticing $g_A=\frac53g_A^q$ \cite{Jiang:2022gjy} and $f_1^{(1)}=\frac{1}{\sqrt2}g_{\pi N\Delta}$ \cite{Jiang:2018mzd}, one has $g_{\pi N\Delta}=\frac{3\sqrt{2}}{5}k_1g_A$. When $k_1=1$, the relation is consistent with Refs. \cite{Hemmert:1997ye,Scherer:2012zzc}. However, the resulting $g_{\pi N\Delta}/g_{\pi NN}$ is not consistent with experimental data \cite{Hemmert:1997ye,Hemmert:1994ky}. A different factor $k_1=\frac54$ from the $SU(4)$ coupling constant relation was adopted in Ref. \cite{Bernard:1995dp}. If one takes the notation of Ref. \cite{Hemmert:1994ky} and compares the ratios there $g^{QM}_{\pi N\Delta}/g^{QM}_{\pi NN}\approx1.7$ and $g_{\pi N\Delta}^{expt}/g_{\pi NN}^{expt}\approx 2.21$, the difference is a number about 1.3. Obviously, $k_1=\frac54$ is a more appropriate value.

Let us consider the multiplication factor $k_1$ in a phenomenological perspective. When we compare the matrix element at the hadron level and the quark level for the same coupling vertex, the correspondence: $\vec{S}_t\rightarrow\vec{\sigma}$ ($k_1=1$) is actually an assumption. In the $\pi NN$ case \cite{Jiang:2022gjy}, the spin operators for hadrons and quarks are both described by the Pauli matrix and the correspondence reads: $\vec{\sigma}_{hadron}\leftrightarrow\vec{\sigma}_{quark}$. The factor 1 is a natural choice. In the present case, however, the hadron-level spin operator differs from the quark-level spin operator. The equivalence of the hadron and quark descriptions can rely on a factor not equal to 1. Here we propose to choose a value $k_1=\sqrt{3/2}$ which is from the third component of the transition operator in Eq. \eqref{St}. This value is numerically close to $k_1=\frac54$ and it gives a $g_{\pi N\Delta}/g_{\pi NN}$ consistent with the experimental data. This choice means that the operator $\sqrt{\frac32}S_t^3$ plays a similar role to the operator $\sigma^3_{hadron}$,
\begin{eqnarray}
|\frac12,\frac12\rangle&=&\sigma^3_{hadron}|\frac12,\frac12\rangle,\nonumber\\
|\frac12,\frac12\rangle&=&\sqrt{\frac32}S_t^3|\frac32,\frac12\rangle.
\end{eqnarray}
The former indicates that $\sigma^3_{hadron}$ transits a state $|\frac12,\frac12\rangle$ into the same state while the latter indicates that $\sqrt{\frac32}S_t^3$ transits a state $|\frac32,\frac12\rangle$ into a state $|\frac12,\frac12\rangle$.

For the value of $k_2$ in the correspondence \eqref{cor2}, one may get $k_2=k_1$ according to the above phenomenological argument. The nonrelativistic reduction for $\sigma_t^{\mu\nu}$ ($\sigma^{\mu\nu}$) is related to $S_t^3$ ($\sigma^3$) when $\mu=1$ and $\nu=2$.

We do not set factors in the correspondences \eqref{cor3}--\eqref{cor5}. One checks these correspondences from theoretical arguments.  All the resulting relations $\hat{e}_i^{(n)}=-3\beta_i^{(n)}$ (also $m_\Delta\approx 3m_q$) are understandable since the spin-1 diquark spectator inside the spin-3/2 $\Delta$ baryon does not affect the behavior of the spin-1/2 operator when it acts on a quark. This is different from the spin-1/2 baryon case \cite{Jiang:2022gjy,Drechsel:1983hny}. Alternatively, the transition of a spin-3/2 state to the same state can always be described by the transition of a spin-1/2 to the same state at quark level. In this situation, the operator $\sigma_{RS}^3$ ($\sigma^3$) transits a state $|\frac32,\frac32\rangle$ ($|\frac12,\frac12\rangle$) into the same state,
\begin{eqnarray}
|\frac32,\frac32\rangle&=&\sigma^3_{RS}|\frac32,\frac32\rangle,\nonumber\\
|\frac12,\frac12\rangle&=&\sigma^3|\frac12,\frac12\rangle.
\end{eqnarray}

When one sets the correspondences \eqref{cor6}--\eqref{cor10}, the necessity of the new factors $k_3$, $\cdots$, $k_7$ can be understood by studying the correspondences \eqref{cor3}--\eqref{cor5} again. If one replaces $k_1S_t^\rho$ with $\gamma^\rho\gamma^5$ in $\bar{S}_{t\rho}S_t^\rho$, the property $\gamma^\mu\gamma_\mu=4$ leads to the requirement of an additional factor to get the correspondence \eqref{cor3}. This factor is different from those in the cases of \eqref{cor4} and \eqref{cor5} by using the same manipulation technique. Therefore, one needs to introduce new factors if the setting procedure for the correspondences with the replacement $k_1S_t^\mu\to\gamma^\mu\gamma^5$ involves the properties of $\gamma$-matrices. Since the meanings of the operators on the left hand sides of \eqref{cor6}--\eqref{cor10} are not clear, one could not determine a value for the $k_3$--$k_7$ at present.

\section{Discussions}\label{sec6}

Relating LECs in different systems from theoretical considerations can reduce the number of independent parameters. In the above sections, we related the LECs in ${\cal L}_{\pi\Delta\Delta}$ and ${\cal L}_{\pi N\Delta}$ to the coupling parameters in the chiral quark model based on the operator correspondences \eqref{cor1}--\eqref{cor10}. The relations between LECs in ${\cal L}_{\pi\Delta\Delta}$ and ${\cal L}_{\pi N\Delta}$ are then easy to get. These hadron-level LECs can be further related to the LECs in ${\cal L}_{\pi NN}$ with the results obtained in Refs. \cite{Jiang:2022gjy,Drechsel:1983hny}. Since the descriptions of spin-3/2 and spin1/2 fields have different formalisms, we introduce several $k$ factors in the correspondences. This is equivalent to modify the values of relevant LECs in $\Delta\chi$PT. Of the introduced seven factors, the $\pi\Delta\Delta$ Lagrangian involves three and the $\pi N\Delta$ Lagrangian involves four. Only two factors ($k_1$ and $k_2$) appearing in ${\cal L}_{\pi N\Delta}$ can be phenomenologically determined at present. The values of the factors $k_3$--$k_7$ may be analyzed in a further study, like the discussions conducted in Ref. \cite{Fettes:2000bb}.

The hadron- and quark-level LEC relations determined with the correspondences \eqref{cor1}--\eqref{cor5} are very simple. A factor of 2 appears in the relations from the correspondences \eqref{cor1} and \eqref{cor2} and a simple factor 3 appears in the relations from the correspondences \eqref{cor3}, \eqref{cor4}, and \eqref{cor5}. Those relations between hadron- and quark-level LECs according to the correspondences \eqref{cor6}--\eqref{cor10} exhibit different forms. The feature in the present case is distinct from that in the spin-1/2 case \cite{Jiang:2022gjy,Drechsel:1983hny}.

In the final results (Table \ref{LECrelations}), many hadron-level LECs vanish. This, however, should not be the realistic case. If one believes that all the hadron-level interactions have quark-level descriptions, such LECs probably get contributions from the diagrams sketched in Fig. \ref{appcor} or from more complicated diagrams. A large number of vanishing LECs indicates the importance of corrections in the case with more Lorentz structures than the spin-1/2 case when we estimate the LECs in the chiral quark model. This issue can be explored in a future work. In higher order Lagrangians, the additional structure $\bar{S}_t^\rho\sigma^{\mu\nu}S^\lambda$ may further complicate the quark model estimation for the LECs. Then corrections would become more essential.

If one adopts the obtained relations in Table \ref{LECrelations}, the number of parameters needs to be fitted with experimental data may be reduced. Although the reduction involving undetermined $k$'s is weaker than the spin-1/2 case, the results are still helpful to future $\Delta$-related studies. Since there are plenty of data in the nucleon-nucleon sector, this sector can be a good platform to justify the present work by including the $\Delta$ fields as intermediate states in the relevant processes.

\section*{Acknowledgments}

This project was supported by the National Natural Science Foundation of China under Grant Nos. 11775132, 11875179, 11905112, 12235008, 12275157, and Guangxi Science Foundation under Grants No. 2022GXNSFAA035489.

%
%
%
%
%

\end{document}